\documentclass[oldversion,letter]{aa}
\usepackage{natbib}
\usepackage{graphicx}
\usepackage{txfonts}

\newcommand{\msunyr}{\ensuremath{\mathit{M}_{\odot}{\rm yr}^{-1}}}   
\newcommand{\kms}{\ensuremath{{\rm km\,s^{-1}}}}                   
\newcommand{\msun}{\ensuremath{\mathit{M}_{\odot}}}   
\newcommand{\lsun}{\ensuremath{\mathit{L}_{\odot}}}                  
\newcommand{\rsun}{\ensuremath{\mathit{R}_{\odot}}}                  

\newcommand{\lstar}{\ensuremath{\mathit{L}_{\star}}}                 
\newcommand{\mdot}{\ensuremath{\dot{M}}}                             
\newcommand{\rstar}{\ensuremath{\mathit{R}_{\star}}}                 
\newcommand{\teff}{\ensuremath{\mathit{T}_{\rm eff}}}                
\newcommand{\reff}{\ensuremath{\mathit{R}_{\rm phot}}}                
\newcommand{\vinf}{\ensuremath{v_{\infty}}}                          
\newcommand{\tstar}{\ensuremath{\mathit{T}_{\star}}}                 

\begin{document}

\title{Massive star evolution: Luminous Blue Variables as unexpected Supernova progenitors}

\author{Jose H. Groh,  Georges Meynet, and Sylvia Ekstr\"om }

\institute{Geneva Observatory, Geneva University, Chemin des Maillettes 51, CH-1290 Sauverny, Switzerland; \email{jose.groh@unige.ch}}

\authorrunning{Groh, Meynet \& Ekstr\"om}
\titlerunning{LBVs as unexpected Supernova progenitors}

\date{Received  / Accepted }

\abstract{Stars more massive than about 8~\msun\ end their lives as a Supernova (SN), an event of fundamental importance Universe-wide. Theoretically, these stars have been expected to be either at the red supergiant, blue supergiant, or Wolf-Rayet stage before the explosion. We performed coupled stellar evolution and atmospheric modeling of stars with initial masses between 20~\msun\ and 120~\msun. We found that the 20~\msun\ and 25~\msun\ rotating models, before exploding as SN, have spectra that do not resemble any of the aforementioned classes of massive stars. Rather, they have remarkable similarities with rare, unstable massive stars known as Luminous Blue Variables (LBV). While observations show that some SNe seem to have had LBVs as progenitors, no theoretical model had yet predicted that a star could explode at this stage. Our models provide theoretical support for relatively low-luminosity LBVs exploding as SN in the framework of single stellar evolution. This is a significant shift in paradigm, meaning that a fraction of LBVs could be the end stage of massive star evolution, rather than a transitory evolutionary phase. We suggest that type IIb SN could have LBV as progenitors, and a prime example could be SN 2008ax.}
\maketitle

\section{\label{intro}Introduction} 

Stars more massive than about eight times the mass of the Sun, although rare, have been key players in the cosmic history through their enormous input of ionizing photons, energy, momentum, and chemical species into the Universe. Their short lives end with a core-collapse Supernova (SN), a generally luminous event that can be traced up to cosmological distances. 

The morphological appearance of massive stars before the SN event is very uncertain. Observationally, red supergiants (RSG) have been confirmed as SN progenitors for stars with up to 18~\msun, while Wolf-Rayet (WR) stars have not yet been confirmed as SN progenitors \citep{smartt09}. Theoretically, progenitors of core-collapse SN have been typically related either to the class of WR, blue supergiant (BSG), or RSG stars \citep{georgy12a,langer12}. Luminous Blue Variables (LBV), a heterogeneous class of unstable massive stars, have also been observationally connected to SN progenitors. This link has been made based on the inference of wind variability from the radio lightcurve \citep{kv06} and line profiles (\citealt{trundle08}; \citealt{gv11}), presence of a dense circumstellar medium \citep{smith07}, and photometry of the progenitor \citep{galyam09,mauerhan12}. However, this has not been supported by stellar evolution theory yet, which had instead suggested LBVs to be in a transitory phase from being an O-type star burning hydrogen in its core to becoming a He-core burning WR star \citep{maeder_araa00,ekstrom12}. Such a glaring discrepancy between observations and theory exposes a main gap in the understanding of massive stars, their deaths, and consequently their impact in the Universe. 

So far, stellar evolution models have been able to self-consistently predict the stellar parameters such as temperature (\tstar) and luminosity (\lstar) only up to the stellar hydrostatic surface. However, especially at the pre-SN stage, this layer is not reached directly by the observations because of the presence of an atmosphere and dense stellar wind. To properly classify the SN progenitor from stellar evolution models, we present here a novel approach, coupling radiative transfer atmospheric/wind calculations done with CMFGEN \citep{hm98} with those performed with the Geneva stellar evolution code \citep{ekstrom12}. Using as input the stellar structure conditions, we are able to generate a synthetic spectrum that can be directly confronted to the observations.  As we discuss in this Letter, having a spectrum allows us to gain enormous insight into the nature of the progenitor. Here we report a quite surprising theoretical finding that single stars with initial mass in the range 20--25~\msun\ have spectra similar to LBVs before exploding as SNe.

\section{\label{model}Physics of the models}

We performed coupled stellar evolution and atmospheric modeling of stars with initial masses between 20~\msun\ and 120~\msun. Hereafter we focus on the results of the 20~\msun\ and 25~\msun\ rotating models. The analysis of the full range of masses will be reported elsewhere (Groh et al. 2013, in preparation).

The evolutionary models are those from \citet{ekstrom12}, computed with the Geneva stellar evolution code. They assume solar metallicity and initial rotational speed of 40\% of the critical velocity for break-up. The prescription for the rotational diffusion coefficients is taken from \citet{zahn92} and \citet{maeder97}. The radiative mass loss rates are implemented following the prescription from \citet{vink01}, and for the RSG phase from \citet{dejager88}. Because of variations in the ionization level of hydrogen beneath the surface of the star during the RSG phase, some layers might exceed the Eddington limit, possibly driving instabilities. In this case, the radiative mass loss is increased by a factor of three in our models, which matches the observations from \citet{vanloon05}.

To compute the output spectra we used the spherically symmetric, full line blanketed, non-local thermodynamical equilibrium, atmospheric radiative transfer code CMFGEN \citep{hm98}.   For the hydrodynamics of the wind, we assume a standard $\beta$-type law with $\beta = 1$, while a hydrostatic solution is computed for the subsonic portion of the atmosphere. The two solutions merge at 0.75 of the sonic speed. The wind terminal velocity is computed using the scaling with the escape velocity from \citet{kudritzki00}. Wind clumping is included with a volume filling factor ($f$) approach, and all models computed here assume $ f = 0.1$.

We use the outputs from the Geneva stellar structure calculations, such as the radius, luminosity, mass, and surface abundances, as input in CMFGEN. For consistency, we adopt the same mass-loss rate recipe as that used by the Geneva evolution code (from \citealt{vink01} in the parameter range analyzed here). We use the temperature structure of the stellar envelope to merge the CMFGEN and the Geneva stellar structure solutions.  Changing the location of the merging point between the CMFGEN and Geneva solutions would change the temperature at the hydrostatic surface, which would make the SN progenitor look hotter or colder. However, the conclusions reached here are independent of this approach. For reasonable changes in the merging layer, no significant difference in spectral morphology is obtained, with the pre-SN model spectra still resembling LBVs. 

\section{\label{fate} The fate of rotating 20--25~\msun\ stars: LBVs exploding as Type II SN}

\begin{figure}
\center
\resizebox{0.995\hsize}{!}{\includegraphics{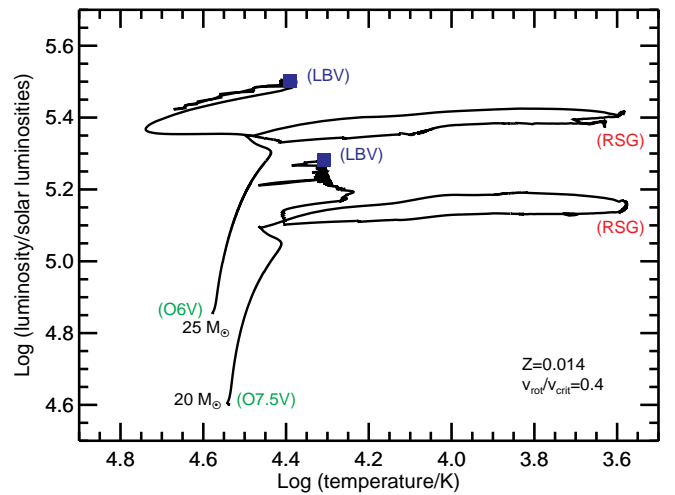}}
\caption{\label{hrd} { Evolutionary tracks of rotating models with initial masses of 20~\msun\ and 25~\msun\ at solar metallicity (Z=0.014), shown in the HR diagram and calculated in \citet{ekstrom12}. The initial rotation rate is 40\% of the critical velocity for break-up. Blue filled squares indicate the pre-SN stage, and representative evolutionary phases are indicated. Temperatures refer to those at the last envelope layer. }}
\end{figure}

Figure~\ref{hrd} shows the evolutionary tracks of rotating massive stars with initial masses of 20~\msun\ and 25~\msun\ at solar metallicity. At the pre-SN stage, the 20~\msun\ model has $\tstar =20\,358$~K and $\lstar= 1.9 \times 10^5$~\lsun\, while the 25~\msun\  has  $\tstar =24\,625$~K and $\lstar= 3.2 \times 10^5$~\lsun\ (blue filled squares in Fig.\ref{hrd}). Several classes of massive stars, such as BSG and LBVs, are found in this range of \tstar\ and \lstar. In the absence of a spectrum, conventional wisdom would have suggested a BSG (based on \tstar\ and \lstar)  or a late-type WN classification (WNL, using the chemical abundance criteria from \citealt{georgy12}).

\begin{figure*}
\center
\resizebox{0.895\hsize}{!}{\includegraphics{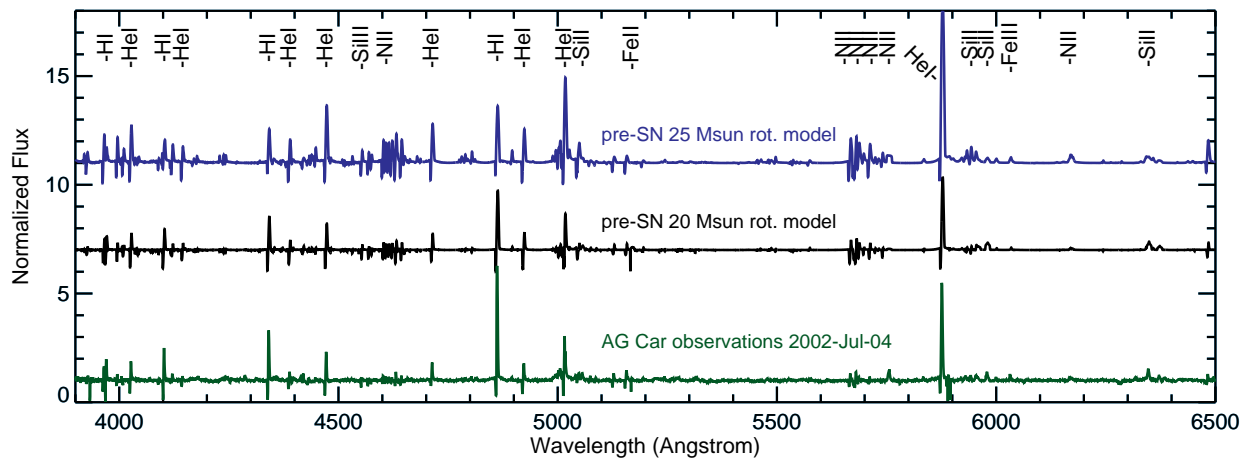}}
\caption{\label{spec} { Synthetic optical spectra of the 20~\msun\ (black) and 25~\msun\ (blue) rotating models at the pre-SN stage. The flux has been continuum-normalized and offset for clarity. The strongest spectral features are identified. We also show the observed spectrum of the prototypical LBV AG Carinae obtained in 2002 July 04 (green),  when the star had an effective temperature of 16400 K \citep{ghd09}. Note that AG Car has a higher luminosity than the pre-SN models, and is shown only to illustrate the morphology of a typical LBV spectrum. We note the remarkable similarity between our synthetic spectra and a typical LBV spectrum. This suggests that some LBV stars are caught just before the explosion. }}
\end{figure*}

Strikingly, we found that the pre-SN synthetic spectra of our models are remarkably similar to the observed spectra of LBVs, such as AG Carinae, HR Carinae, P Cygni, HDE 316285, among others \citep{ghd09,gdh09,najarro01,hillier98}. Our results show that the spectral morphology of the 20 \msun\  and 25~\msun\ pre-SN models is markedly different than that of RSG, BSG, or WR stars. Figure~\ref{spec}  shows the pre-SN synthetic optical spectra computed with our atmospheric modeling. To illustrate the morphology of a  typical observed LBV spectrum, we also display the prototypical LBV AG Carinae in Fig.~\ref{spec}. 

The similarities are conspicuous, especially because our models do not have initial mass, rotation rate, mass-loss rate, or wind terminal velocity tuned to reproduce the spectrum of AG Car or any other LBV. One can clearly see that the main features present in the optical spectrum of LBVs, such as strong H, \ion{He}{i}, and \ion{N}{ii} lines with P-Cygni profiles \citep{ghd09,stahl93}, are also present in the synthetic spectra at the pre-SN stage.  While some LBVs may display a WN11h spectral type at visual minimum, we note that \ion{He}{ii} $\lambda4686$ is not present in emission and, therefore, a WR classification with the latest possible spectral types (WN9h - WN11h) would be inappropriate \citep{crowther95a}. We obtained that the morphological appearance of the SN progenitor is regulated not only by \tstar\ and \lstar\ but also by the stellar wind properties, such as the mass-loss rate and wind terminal velocity. In particular, the modeled stars have relatively high mass-loss rate ($\mdot \sim 1-4\times10^{-5}\msunyr$) and low wind terminal velocity ($\vinf \sim270$ to $320~\kms$), which are in line with values typically encountered in LBVs. The model properties at the pre-SN stage are summarized in Table~\ref{params}.

The Galactic LBVs span a range of luminosities, with $\log (\lstar/\lsun)\simeq 5.2 - 6.5$ \citep{clark05}. Observationally, it is unclear whether low- and high-luminosity LBVs represent well-defined, separated classes. Our results seem to indicate that low-luminosity LBVs ($\log (\lstar/\lsun)\simeq 5.2 - 5.5$) are the endpoints of stellar evolution, while high-luminosity LBVs could still evolve to the WR phase.

In addition to spectroscopic similarities, both the 20~\msun\ and 25~\msun\ rotating models are extremely close to the Eddington limit, a limit at which the luminosity of the star is so high that the gravity could not longer hold the star in hydrostatic equilibrium. Our modeling indicates that the Eddington parameter $\Gamma$ is around 0.85 -- 0.90 in the deep subsonic layers of the wind (at a Rosseland optical depth of 20). This occurs because of the relatively high luminosity-to-mass ratio resulting from the evolution, which in turn supports the prescribed mass-loss rates and wind-terminal velocities used in the stellar evolution and spectral calculations. We recall that the proximity to (or surpassing of) the Eddington limit seems to be a key factor in the LBV phenomenon \citep{hd94,so06,ghd11,grafener12a}. Our assumed mass-loss rates may be affected by the distinct properties of stellar mass loss near the Eddington limit \citep{grafener08,vink11}, which could change the extension of the atmosphere and the effective temperature. However, for reasonable changes, qualitatively we expect similar results, i.\,e., a spectral type reminiscent of LBVs would still be seen.

A third kinship between our models and LBVs are the chemical abundances at the surface. Both models show He and N overabundances and H, C, and O depletion, which is a signature that CNO-processed material is present on the surface (Fig. \ref{abund}). For the 20~\msun\ model, the He and N abundances (0.74 and 0.0075 by mass fraction, respectively) are remarkably similar to those of LBVs \citep[e.\,g.][]{clark09}. The 25~\msun\ model predicts a more chemically enriched surface with He and N (0.92 and 0.016 by mass fraction, respectively) and, as expected, its spectrum shows stronger He I and N II lines than that of the 20~\msun\ model or the LBV observations. Still, the 25~\msun\ model spectrum looks more similar to LBVs than to any other class of massive stars.

We note that the LBV classification is phenomenological, where one of the following properties has to be present. Either the star shows S-Doradus type variability, or giant eruptions of several solar masses must have occurred \citep{hd94,vg01}. Therefore, had the SN model progenitors been observed, they would technically be classified as ``candidate LBVs", since none of the above phenomena are included in our calculations. Despite semantics, here we show that the model SN progenitors are alike stars in the quite rare LBV phase in many aspects, such as spectral appearance, proximity to the Eddington limit, and chemical abundances at the surface.

One may wonder why the rotating model explodes as an LBV, while the corresponding non-rotating model 20~\msun\ does not evolve back to the blue and explodes as a RSG, showing a type IIP SN. This is mainly an effect of rotational mixing which produces a larger He-core. The He-core is the inner part of the star, enclosed in the first layer encountered going deeper into the star where the He mass fraction is greater than $\sim0.75$. It is a well known effect that the larger is the mass fraction of the total mass occupied by the He-core, the bluer is the position of the star in the HR diagram  \citep{giannone67}. In our rotating 20~\msun\ model, the He-core encompasses 100\% of the total mass, while in the non-rotating corresponding model, it encompasses only 71\%. Hence the difference of position in the HR diagram. 

\section{Implications for massive star evolution}

Our results indicate that for rotating massive stars with initial masses between 20--25~\msun, a previously unreported evolutionary sequence occurs. The star appears at the Zero-Age Main Sequence as late-type O  star (O6V to O7.5V), evolving to a BSG/O supergiant (OSG), a RSG,  a BSG/blue hypergiant (BHG, for the 20~\msun\ model) or O supergiant (OSG)/WNL  (for the 25~\msun\ model), and finally appearing as an LBV before the SN event,

{\it 20~\msun:  O7.5V $\rightarrow$ BSG $\rightarrow$  RSG  $\rightarrow$ BSG/BHG $\rightarrow$ LBV $\rightarrow$ SN;}

{\it 25~\msun:  O6V $\rightarrow$ OSG $\rightarrow$  RSG  $\rightarrow$ OSG/WNL $\rightarrow$ LBV $\rightarrow$ SN.}\\
Our models indicate that the pre-mortem LBV phase is short ($\sim5000$ years), which is consistent with the observed rarity of LBVs \citep{hd94,vg01,clark05}. The 20~\msun\  and 25~\msun\ models lose most of their mass at the RSG phase (8.7~\msun\ and 9.6~\msun\, respectively), which enables the star to have the aforementioned surface properties required to show an LBV spectrum at pre-death. Thus, mass loss at the RSG phase has a key impact in the subsequent evolution (see \citealt{georgy12}).

\begin{table*}[!ht]
\begin{minipage}{\textwidth}
\caption{Properties of the SN progenitors from rotating models at solar metalicity, which have an LBV spectrum and explode as SN II-b / II-n.}
\label{params}
\centering
\scalebox{0.9}{
\begin{tabular}{l c c c c c c c c c c  c c}
\hline\hline
M$_\mathrm{ini}$& M$_\mathrm{prog.}$\tablefootmark{a}& Age\tablefootmark{a} & $\lstar$\tablefootmark{a} & $\tstar$\tablefootmark{a,b} & $\teff$\tablefootmark{c} & $\rstar$\tablefootmark{a,b} & $\reff$\tablefootmark{c} & $\mdot$\tablefootmark{a} & $\vinf$  & $M_V$ &$B-V$ &He (surface) \\
(\msun) & (\msun) & (Myr) & ($10^5~\lsun$) & (K) & (K) & (\rsun) & (\rsun) & (\msunyr) & (\kms) & (mag) & (mag) & (mass fraction) \\  
\hline
 20.0   & 7.1   & 10.5 & $1.9$ &  20355 &  19540 & 35.3 & 38.1 & $1.2\times10^{-5}$ &   272 & $-6.8$ & $-0.14$ &  0.74 \\
 25.0  & 9.6  & 8.6 & $3.2$ &  25790 &  20000 & 28.4 & 47.0 &  $4.6\times10^{-5}$ &   326  & $-7.2$ & $-0.11$ &  0.92 \\
\hline
\end{tabular}}
\tablefoot{\tablefoottext{a}{From \citet{ekstrom12}.}\tablefoottext{b}{Defined here as the temperature (radius) of the last layer of the stellar envelope.}\tablefoottext{c}{Defined here as the temperature (radius) of the layer where the Rosseland optical depth is equal to 2/3.}}
\end{minipage}
\end{table*}

\begin{figure}
\center
\resizebox{0.895\hsize}{!}{\includegraphics{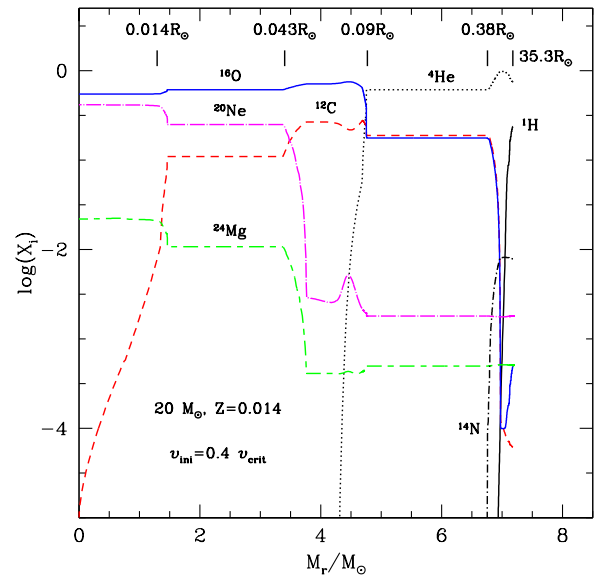}}
\caption{\label{abund} {Variation of the abundances in mass fraction inside our 20~\msun\ model at the end of the core carbon-burning stage. At the top, we indicate the radius in solar units of selected shells inside the star.}}
\end{figure}

Let us look at the expected properties of the SN arising from the explosion of these low-luminosity LBVs. For this purpose, we investigate the chemical composition of the star before the explosion. Figure~\ref{abund} shows the chemical structure of our 20~\msun\  rotating model at the end of the core C burning phase. The star consists of a He-rich core containing  94\% of the total mass and extending over slightly more than 10\% of the total radius of the star, surrounded by a very shallow H-rich envelope containing about 6\% of the mass and extending over about 90\% of the total radius. We see that very little amount of H remains at the surface.

According to \citet{georgy12a}, the 20~\msun\  model presents a very nice fit with the observed characteristics of the core-collapse SN 2008ax \citep{crockett08}. These are (a) a CO core mass between 4 and 5~\msun\ (the model gives 4.73~\msun), (b) a small amount of hydrogen in the ejecta (the model gives 0.02~\msun\  of H) and (c) the estimated position of the progenitor in the color-magnitude diagram (Fig. 10 of \citealt{georgy12a}). Our revised colors for the rotating 20~\msun\  model ($B-V=-0.14$) confirms the very good agreement. SN 2008ax was found to be a type IIb, i.e. an event which transitions from type II to Ib. This suggests that type IIb SN could have an LBV as progenitor, and a prime example could be SN 2008ax.
 
Thus, for the rotating 20~\msun\ model, one would expect that the SN progenitor, which has an LBV-type spectrum, will explode as a type IIb SN. The 25 M$_\odot$ model presents a very similar structure as the one shown in Fig.~\ref{abund}, thus will likely explode as a type IIb SN too. Because the progenitor star has significant mass loss during the RSG phase and crosses the infamous ``Yellow Void" in the HR diagram \citep{dejager98}, enhanced mass loss may have occurred in both phases. As a consequence, this would produce a H-rich circumstellar medium with dense, slow velocity material, and its interaction with the SN blast wave could produce a type IIn SN instead of a type IIb. 

The (still scarce) observational evidence that suggests that LBVs can be SN progenitors indicates that the progenitors could come from a range of initial masses. Some of the LBVs suggested as SN progenitors had their initial mass poorly constrained, such as SN 2005gj \citep{trundle08}, while others were linked to very massive stars ($>50~\msun$), such as SN 2005gl \citep{galyam09} and SN 2006gy \citep{smith07}. Interestingly, some SNe have been proposed to have LBVs of modest initial masses as progenitors, comparable to the models we compute here, as in the case of SN 1987A \citep{smith07b}.

While our models still do not support LBVs as progenitors of SNe from stars with initial mass above ~25~\msun, they provide unexpected theoretical support for explosions of low-luminosity LBVs in a single star scenario. The number of low-luminosity LBVs that explode as SN depends on the exact mass limits and distribution of rotational velocities at birth, since the non-rotating models explode as a RSG (20~\msun) or WNL/LBV (25~\msun). We note that the time average rotational velocities of the 20 and 25~\msun\ models during the Main Sequence (MS) are 217 and 209 km s$^{-1}$. This is comparable to the usual 200 km s$^{-1}$ considered as the mean velocities of such stars during the MS phase. Thus, one would expect that a significant fraction of stars in this mass range would follow the evolution previously discussed and explode as an LBV.

In summary, our key theoretical finding is that stars with spectrum similar to LBVs immediately precede core collapse and the subsequent SN event of rotating stars with initial mass between 20~\msun\ and 25~\msun. Our models predict that the low-luminosity LBV at the pre-SN stage are blue, with $B-V=-0.14$ to $-0.11$. They also have absolute magnitudes in the V-band between $M_V = -6.8$ and $-7.2$. These values are three magnitudes fainter than those of SN progenitors that have been proposed to be LBVs, as in SN 2005gl and 2009ip ($M_V \sim -10$; \citealt{galyam09,mauerhan12}). Given the obvious observational bias towards detecting the more luminous progenitors, a population of SNe with low-luminosity LBVs as progenitors could be awaiting detection, and the previous evidence for LBVs being SN progenitors could have revealed just the tip of the iceberg. 

\begin{acknowledgements}

JHG is supported by an Ambizione Fellowship of the Swiss National Science Foundation. We thank the referee, Goetz Graefener, for detailed comments and John Hillier for making CMFGEN available.

\end{acknowledgements}
\bibliography{../../refs}

\begin{thebibliography}{38}
\expandafter\ifx\csname natexlab\endcsname\relax\def\natexlab#1{#1}\fi

\bibitem[{{Clark} {et~al.}(2009){Clark}, {Crowther}, {Larionov}, {Steele},
  {Ritchie}, \& {Arkharov}}]{clark09}
{Clark}, J.~S., {Crowther}, P.~A., {Larionov}, V.~M., {et~al.} 2009, \aap, 507,
  1555

\bibitem[{{Clark} {et~al.}(2005){Clark}, {Larionov}, \& {Arkharov}}]{clark05}
{Clark}, J.~S., {Larionov}, V.~M., \& {Arkharov}, A. 2005, \aap, 435, 239

\bibitem[{{Crockett} {et~al.}(2008){Crockett}, {Eldridge}, {Smartt},
  {Pastorello}, {Gal-Yam}, {Fox}, {Leonard}, {Kasliwal}, {Mattila}, {Maund},
  {Stephens}, \& {Danziger}}]{crockett08}
{Crockett}, R.~M., {Eldridge}, J.~J., {Smartt}, S.~J., {et~al.} 2008, \mnras,
  391, L5

\bibitem[{{Crowther} {et~al.}(1995){Crowther}, {Hillier}, \&
  {Smith}}]{crowther95a}
{Crowther}, P.~A., {Hillier}, D.~J., \& {Smith}, L.~J. 1995, \aap, 293, 172

\bibitem[{{de Jager}(1998)}]{dejager98}
{de Jager}, C. 1998, \aapr, 8, 145

\bibitem[{{de Jager} {et~al.}(1988){de Jager}, {Nieuwenhuijzen}, \& {van der
  Hucht}}]{dejager88}
{de Jager}, C., {Nieuwenhuijzen}, H., \& {van der Hucht}, K.~A. 1988, \aaps,
  72, 259

\bibitem[{{Ekstr{\"o}m} {et~al.}(2012){Ekstr{\"o}m}, {Georgy}, {Eggenberger},
  {Meynet}, {Mowlavi}, {Wyttenbach}, {Granada}, {Decressin}, {Hirschi},
  {Frischknecht}, {Charbonnel}, \& {Maeder}}]{ekstrom12}
{Ekstr{\"o}m}, S., {Georgy}, C., {Eggenberger}, P., {et~al.} 2012, \aap, 537,
  A146

\bibitem[{{Gal-Yam} \& {Leonard}(2009)}]{galyam09}
{Gal-Yam}, A. \& {Leonard}, D.~C. 2009, \nat, 458, 865

\bibitem[{{Georgy}(2012)}]{georgy12}
{Georgy}, C. 2012, \aap, 538, L8

\bibitem[{{Georgy} {et~al.}(2012){Georgy}, {Ekstr{\"o}m}, {Meynet}, {Massey},
  {Levesque}, {Hirschi}, {Eggenberger}, \& {Maeder}}]{georgy12a}
{Georgy}, C., {Ekstr{\"o}m}, S., {Meynet}, G., {et~al.} 2012, \aap, 542, A29

\bibitem[{{Giannone}(1967)}]{giannone67}
{Giannone}, P. 1967, \zap, 65, 226

\bibitem[{{Gr{\"a}fener} \& {Hamann}(2008)}]{grafener08}
{Gr{\"a}fener}, G. \& {Hamann}, W.-R. 2008, \aap, 482, 945

\bibitem[{{Gr{\"a}fener} {et~al.}(2012){Gr{\"a}fener}, {Owocki}, \&
  {Vink}}]{grafener12a}
{Gr{\"a}fener}, G., {Owocki}, S.~P., \& {Vink}, J.~S. 2012, \aap, 538, A40

\bibitem[{{Groh} {et~al.}(2009{\natexlab{a}}){Groh}, {Damineli}, {Hillier},
  {Barb{\'a}}, {Fern{\'a}ndez-Laj{\'u}s}, {Gamen}, {Mois{\'e}s}, {Solivella},
  \& {Teodoro}}]{gdh09}
{Groh}, J.~H., {Damineli}, A., {Hillier}, D.~J., {et~al.} 2009{\natexlab{a}},
  \apjl, 705, L25

\bibitem[{{Groh} {et~al.}(2011){Groh}, {Hillier}, \& {Damineli}}]{ghd11}
{Groh}, J.~H., {Hillier}, D.~J., \& {Damineli}, A. 2011, \apj, 736, 46

\bibitem[{{Groh} {et~al.}(2009{\natexlab{b}}){Groh}, {Hillier}, {Damineli},
  {Whitelock}, {Marang}, \& {Rossi}}]{ghd09}
{Groh}, J.~H., {Hillier}, D.~J., {Damineli}, A., {et~al.} 2009{\natexlab{b}},
  \apj, 698, 1698

\bibitem[{{Groh} \& {Vink}(2011)}]{gv11}
{Groh}, J.~H. \& {Vink}, J.~S. 2011, \aap, 531, L10

\bibitem[{{Hillier} {et~al.}(1998){Hillier}, {Crowther}, {Najarro}, \&
  {Fullerton}}]{hillier98}
{Hillier}, D.~J., {Crowther}, P.~A., {Najarro}, F., \& {Fullerton}, A.~W. 1998,
  \aap, 340, 483

\bibitem[{{Hillier} \& {Miller}(1998)}]{hm98}
{Hillier}, D.~J. \& {Miller}, D.~L. 1998, \apj, 496, 407

\bibitem[{{Humphreys} \& {Davidson}(1994)}]{hd94}
{Humphreys}, R.~M. \& {Davidson}, K. 1994, \pasp, 106, 1025

\bibitem[{{Kotak} \& {Vink}(2006)}]{kv06}
{Kotak}, R. \& {Vink}, J.~S. 2006, \aap, 460, L5

\bibitem[{{Kudritzki} \& {Puls}(2000)}]{kudritzki00}
{Kudritzki}, R.-P. \& {Puls}, J. 2000, \araa, 38, 613

\bibitem[{{Langer}(2012)}]{langer12}
{Langer}, N. 2012, \araa, 50, 107

\bibitem[{{Maeder}(1997)}]{maeder97}
{Maeder}, A. 1997, \aap, 321, 134

\bibitem[{{Maeder} \& {Meynet}(2000)}]{maeder_araa00}
{Maeder}, A. \& {Meynet}, G. 2000, \araa, 38, 143

\bibitem[{{Mauerhan} {et~al.}(2012){Mauerhan}, {Smith}, {Filippenko},
  {Blanchard}, {Blanchard}, {Casper}, {Cenko}, {Clubb}, {Cohen}, {Li}, \&
  {Silverman}}]{mauerhan12}
{Mauerhan}, J.~C., {Smith}, N., {Filippenko}, A., {et~al.} 2012, ArXiv e-prints

\bibitem[{{Najarro}(2001)}]{najarro01}
{Najarro}, F. 2001, in Astronomical Society of the Pacific Conference Series,
  Vol. 233, P Cygni 2000: 400 Years of Progress, ed. M.~{de Groot} \&
  C.~{Sterken}, 133

\bibitem[{{Smartt}(2009)}]{smartt09}
{Smartt}, S.~J. 2009, \araa, 47, 63

\bibitem[{{Smith}(2007)}]{smith07b}
{Smith}, N. 2007, \aj, 133, 1034

\bibitem[{{Smith} {et~al.}(2007){Smith}, {Li}, {Foley}, {Wheeler}, {Pooley},
  {Chornock}, {Filippenko}, {Silverman}, {Quimby}, {Bloom}, \&
  {Hansen}}]{smith07}
{Smith}, N., {Li}, W., {Foley}, R.~J., {et~al.} 2007, \apj, 666, 1116

\bibitem[{{Smith} \& {Owocki}(2006)}]{so06}
{Smith}, N. \& {Owocki}, S.~P. 2006, \apjl, 645, L45

\bibitem[{{Stahl} {et~al.}(1993){Stahl}, {Mandel}, {Wolf}, {Gaeng}, {Kaufer},
  {Kneer}, {Szeifert}, \& {Zhao}}]{stahl93}
{Stahl}, O., {Mandel}, H., {Wolf}, B., {et~al.} 1993, \aaps, 99, 167

\bibitem[{{Trundle} {et~al.}(2008){Trundle}, {Kotak}, {Vink}, \&
  {Meikle}}]{trundle08}
{Trundle}, C., {Kotak}, R., {Vink}, J.~S., \& {Meikle}, W.~P.~S. 2008, \aap,
  483, L47

\bibitem[{{van Genderen}(2001)}]{vg01}
{van Genderen}, A.~M. 2001, \aap, 366, 508

\bibitem[{{van Loon} {et~al.}(2005){van Loon}, {Cioni}, {Zijlstra}, \&
  {Loup}}]{vanloon05}
{van Loon}, J.~T., {Cioni}, M.-R.~L., {Zijlstra}, A.~A., \& {Loup}, C. 2005,
  \aap, 438, 273

\bibitem[{{Vink} {et~al.}(2001){Vink}, {de Koter}, \& {Lamers}}]{vink01}
{Vink}, J.~S., {de Koter}, A., \& {Lamers}, H.~J.~G.~L.~M. 2001, \aap, 369, 574

\bibitem[{{Vink} {et~al.}(2011){Vink}, {Muijres}, {Anthonisse}, {de Koter},
  {Gr{\"a}fener}, \& {Langer}}]{vink11}
{Vink}, J.~S., {Muijres}, L.~E., {Anthonisse}, B., {et~al.} 2011, \aap, 531,
  A132

\bibitem[{{Zahn}(1992)}]{zahn92}
{Zahn}, J.-P. 1992, \aap, 265, 115

\end{thebibliography}

\end{document}